%% file: main.tex
\documentclass{article}

    \PassOptionsToPackage{numbers, compress}{natbib}
\usepackage[preprint]{neurips_2026}

\usepackage[utf8]{inputenc} 
\usepackage[T1]{fontenc}    
\usepackage{hyperref}       
\hypersetup{
    colorlinks=true,
    linkcolor=black,
    citecolor=black,
    filecolor=black,
    urlcolor=black,
}
\usepackage{url}            
\usepackage{booktabs}       
\usepackage{amsfonts}       
\usepackage{nicefrac}       
\usepackage{microtype}      
\usepackage{xcolor}         
\usepackage{tikz,lipsum,lmodern}
\usepackage[most]{tcolorbox}

\usepackage{subcaption}

\usepackage{array} 

\usepackage{multirow} 

\title{AI Security Research Should Better Incentivize Defense Research}

\author{%
  Youqian Zhang\\
  The Hong Kong Polytechnic University\\
  \texttt{you-qian.zhang@polyu.edu.hk} \\
}

\begin{document}

\maketitle

\begin{abstract}
\input{abstrac}
\end{abstract}

\input{introduction}

\input{methodology}

\input{results}

\input{dilemma_for_defense_research}

\input{alternative_views}

\input{limitations}

\input{conclusion}

\bibliographystyle{plainnat}
\small
\bibliography{main}

\newpage
\appendix
\input{appedix_soks}

\newpage
\input{appendix_surveys}


\end{document}

%% file: abstrac.tex

This work examines an imbalance in artificial intelligence (AI) security research: the field tends to produce more work on attacking AI systems than on defending them. 
Drawing on related academic papers, we find biased attack-to-defense ratios across subfields, including federated learning, speech recognition, membership inference, large language models, etc. 
The imbalance possibly means far beyond a simple count: attack papers are routinely evaluated under favorable conditions that make threats look more severe than they are in practice, while defenses are held to a stricter standard that few can meet. 
The result is a literature rich in demonstrated vulnerabilities and thin on usable and deployed protections. 
We thus argue that AI security research should better incentivize defense research.

%% file: introduction.tex
\section{Introduction}
\label{sec:introduction}

The rapid progress of artificial intelligence (AI) has enabled more and more smart systems across vision, language, healthcare, and decision-making. 
As AI systems are deployed in more critical and consequential settings, concerns about their security and privacy have become urgent. 
Even accurate models can be vulnerable to carefully crafted inputs, data leakage, or manipulation during deployment, raising the risk that AI systems may fail in ways that are harmful, exploitable, and difficult to detect.

These concerns are no longer hypothetical. 
For example, in computer vision, Szegedy et al.,~\cite{szegedy14intriguing} in 2014 demonstrated that neural
networks are vulnerable to adversarial examples.
Small perturbations to a few pixels of an input image can cause a model to misclassify a cat as a dog, despite the change being nearly imperceptible to humans~\cite{su19one}. 
This may not sound that harmful, however, in physical-world settings, such vulnerabilities can take the form of stickers, patterns, or camouflage-like modifications that alter a model’s prediction under real deployment conditions, and in safety-critical applications such as autonomous driving, such failures may have serious consequences (e.g., traffic rule violations or car crashes)~\cite{guesmi2023physical}. 
Privacy risks are similarly significant. 
Modern AI systems are trained on large-scale datasets that often contain sensitive information, including biometric records and medical data. 
This has motivated a wide range of attacks showing that an adversary with access to a model may infer whether a particular example was used in training~\cite{bai2024membership}, extract sensitive attributes~\cite{oliynyk2023know}, or even reconstruct parts of the training data~\cite{salem2023sok}.

A large body of work has emerged to expose weaknesses in AI systems. We refer to this line of work broadly as ``attack research'', which identifies, characterizes, or demonstrates vulnerabilities that can be exploited to compromise the security and privacy of AI. 
The counterpart is ``defense research'', which develops and validates methods for preventing, mitigating, or withstanding these attacks. 
Attack and defense research are both essential: attacks diagnose weaknesses, while defenses are what ultimately make systems secure.

This position paper argues that the current research landscape places disproportionate emphasis on attacks relative to defenses. 
Based on a quantitative analysis of cited studies from 16 systematization-of-knowledge (SoK) papers, covering 1,162 attack/defense-classified citations, we find that attack papers outnumber defense papers by 1.24:1.
In the cited studies of another 21 survey papers, the attack papers outnumber defense papers by 2.15 on average.
We do not claim that attack and defense papers must exist in equal numbers.
Our concern is instead that the present imbalance is observable and consistent enough, to suggest a structural lag in defensive progress. 
In some areas, the literature appears much better at demonstrating how AI systems fail than at developing robust, deployable ways to make them secure. 
This is a problematic signal for a field whose societal impact increasingly depends not only on identifying vulnerabilities, but also on mitigating them.
Our position is as follows: 

\begin{tcolorbox}[colback=black!3!white,colframe=black,title=Position]
  \textbf{AI security research should better
incentivize defense research.}
\end{tcolorbox}

We argue that this is not merely a matter of filling a topical gap, but of correcting an incentive structure that systematically rewards the discovery of vulnerabilities more than their remediation. 
While AI security research is maturing as a field, it needs to invest more deliberately in defenses as first-class scientific contributions.

%% file: methodology.tex
\section{Collecting Attack and Defense Papers}

We began by collecting 16 Systematization of Knowledge (SoK) papers~\cite{khanna2024sok,li2023sok,wang2025sok,noppel2024sok,nayan2024sok,nowmi2025critical,duddu2024sok,papernot2018sok,carletti2025sok,abdullah2021sok,du2025sok,horvath2024sok,wenger2023sok,meeus2025sok,salem2023sok,suya2024sok}\footnote{Two studies~\cite{wang2025sok,nowmi2025critical} have been accepted by IEEE S\&P 2026 but have not yet appeared in official proceedings when this paper is written; for these, we use publicly available preprint versions.} on AI security and privacy from top-tier security and privacy venues, including USENIX Security, IEEE Symposium on Security and Privacy (IEEE S\&P), and IEEE Conference on Secure and Trustworthy Machine Learning (IEEE SaTML). We focus on SoK papers because they primarily synthesize, organize, and contextualize existing research in order to provide a structured understanding of a relatively mature area.
From these 16 SoK papers, we extracted their bibliographies and manually classified cited works into ``Attack'' and ``Defense''\footnote{Annotator(s) first used the paper title and the citation context or description provided in the SoK paper. If the label was still unclear, we consulted the abstract. Papers were labeled Attack if their main contribution was to identify, demonstrate, improve, or evaluate a method for exploiting vulnerabilities; Defense if their main contribution was to prevent, detect, mitigate, or certify against such vulnerabilities; and Other for surveys, benchmarks, datasets, taxonomies, and papers whose contribution was not primarily offensive or protective. When papers contained both attack and defense elements, we assigned the label corresponding to the dominant contribution; if no dominant contribution was clear, we labeled the paper as Other. For concise presentation, we exclude Other from further analysis}. The resulting data were organized into an Excel workbook with one sheet per SoK paper, along with an overview sheet aggregating all entries. For each cited paper, we recorded its manually assigned label, publication year, and venue. We also normalized venue names to common short forms; for example, ``IEEE Symposium on Security and Privacy'', ``Oakland'', and ``S\&P'' were mapped to ``IEEE S\&P''. 
After normalization, venues were grouped into four broader categories for venue-level analysis: ``Security'', ``AI'', ``Preprint'', and ``Other''. Except for preprints, all three other categories consist of peer-reviewed venues.
In addition, to identify papers cited across multiple SoKs, we performed title-based deduplication. Specifically, titles were normalized by converting them to lowercase, removing punctuation, and collapsing repeated whitespace, and the resulting normalized title was used as the deduplication key. After deduplication, the SoK corpus contained 644 unique attack papers and 518 unique defense papers.

To broaden coverage beyond SoK papers, we then incorporated 21 additional survey papers spanning several subareas, including membership inference attacks (MIA)~\cite{hu2022membership,bai2024membership,hu2023defenses,wu2025membership}, adversarial examples~\cite{wei2024physical,nguyen2023physical,guesmi2023physical,wei2026visual,serban2020adversarial}, large language models (LLMs)~\cite{knowlton2026prompt,hakim2026jailbreaking,yi2024jailbreak,dong2025safeguarding}, model stealing~\cite{oliynyk2023know,gencc2023taxonomic}, federated learning (FL)~\cite{zhao2025federation,yang2023gradient}, speech recognition~\cite{tan2022adversarial}, deepfakes~\cite{edwards2024review,pei2024deepfake,li2025survey}, and machine unlearning~\cite{liu2025threats,shaik2024exploring,nguyen2025survey}. For these additional surveys, we manually read each paper and recorded the counts of attack and defense papers that matched the topic of each.

The dataset used in this study is available at: \url{https://osf.io/sdmwk/overview?view_only=be7a5597cfce445e9ec086b62d9ec631}.
Summaries of the SoK papers and the survey papers are in Appendix~\ref{app:soks} and Appendix~\ref{app:surveys}, respectively.

%% file: results.tex
\section{Imbalance Between Attack and Defense Research}

\subsection{Per-SoK Breakdown}

\begin{figure}[t]
    \centering
    \includegraphics[width=0.9\linewidth]{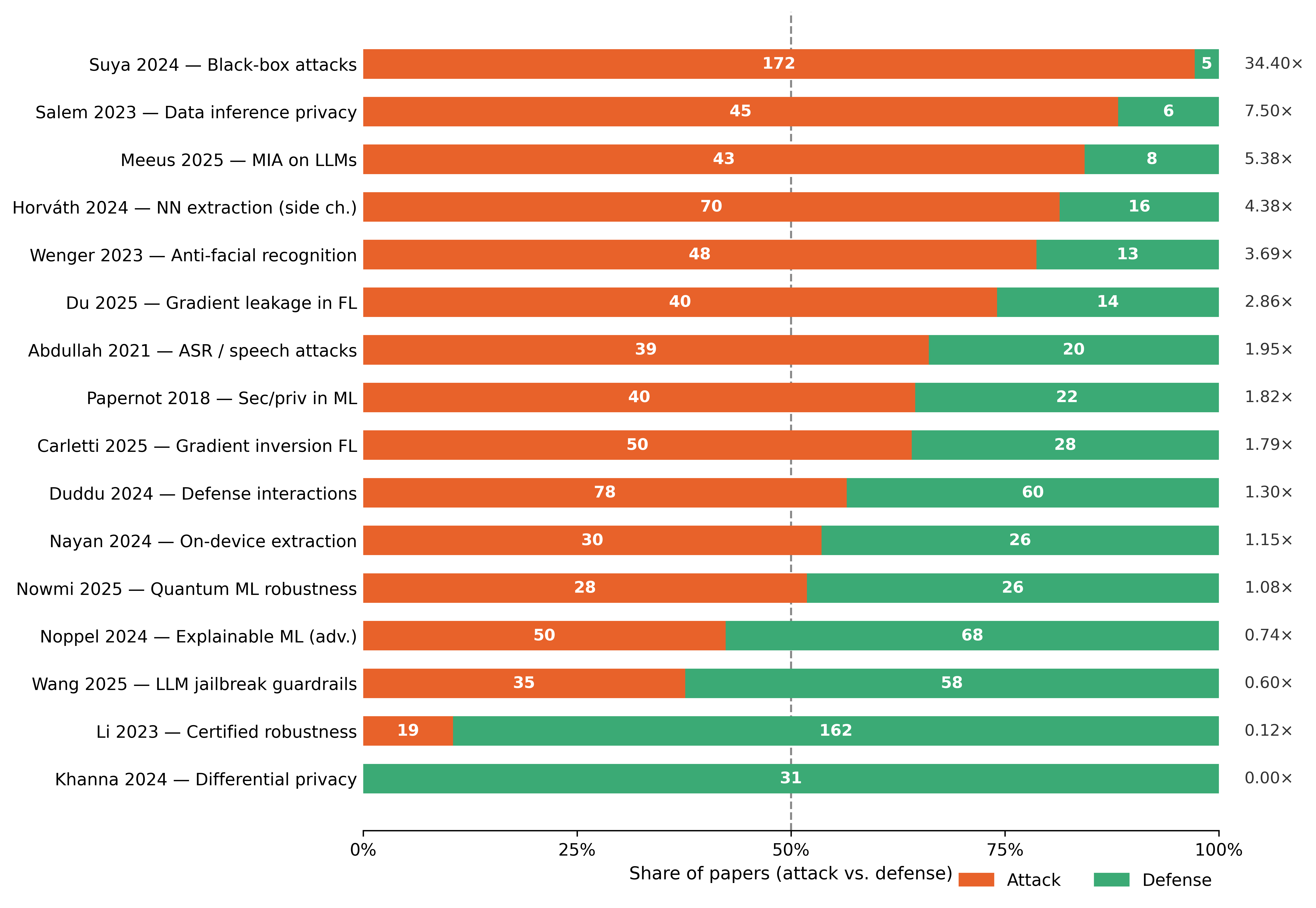}
    \caption{Attack vs. defense paper composition across 16 SoK papers.}
    \label{fig:per_sok_breakdown}
\end{figure}

Disaggregating the cited papers by individual SoK shows that the overall imbalance is not driven by only a few outliers, but is distributed across much of the literature. As shown in Figure~\ref{fig:per_sok_breakdown}, 11 of the 16 SoKs cite more attack papers than defense papers. The attack-to-defense ratio (``A:D''), defined as the ratio of attack-paper counts to defense-paper counts, varies substantially across topics, ranging from near balance in some areas to extreme asymmetry in others. 

A first interesting pattern is that the most attack-heavy SoKs are all organized around a specific attack class. The five most skewed cases---Suya et al.~\cite{suya2024sok} (34.40$\times$), Salem et al.~\cite{salem2023sok} (7.50$\times$), Meeus et al.~\cite{meeus2025sok} (5.38$\times$), Horv\'ath et al.~\cite{horvath2024sok} (4.38$\times$), and Wenger et al.~\cite{wenger2023sok} (3.69$\times$)---focus, respectively, on black-box adversarial attacks, data inference privacy attacks, membership inference against LLMs, neural network extraction via physical side channels, and anti-facial-recognition techniques\footnote{We classify anti-facial-recognition techniques as attack research because they primarily rely on offensive methods designed to evade or defeat facial-recognition systems.}. In these areas, the literature appears to be structured primarily around discovering, characterizing, and refining offensive capabilities, while defensive work tends to emerge later and in a more reactive role rather than as a coequal research agenda.

A complementary pattern appears among the defense-heavy SoKs. Khanna et al.~\cite{khanna2024sok}, Li et al.~\cite{li2023sok}, and Wang et al.~\cite{wang2025sok} review topics whose organizing concepts are themselves defensive mechanisms or assurance frameworks, such as differential privacy, certified robustness, and jailbreak guardrails. 
We noticed that defense-heavy SoKs usually emerged because they focused on a specific defense method, not because defenses are equally prominent in broad surveys of the threat landscape.

\subsection{Venue Distribution}

\begin{figure}[t]
    \centering
    \begin{subfigure}[t]{0.49\linewidth}
        \centering
        \includegraphics[width=\linewidth]{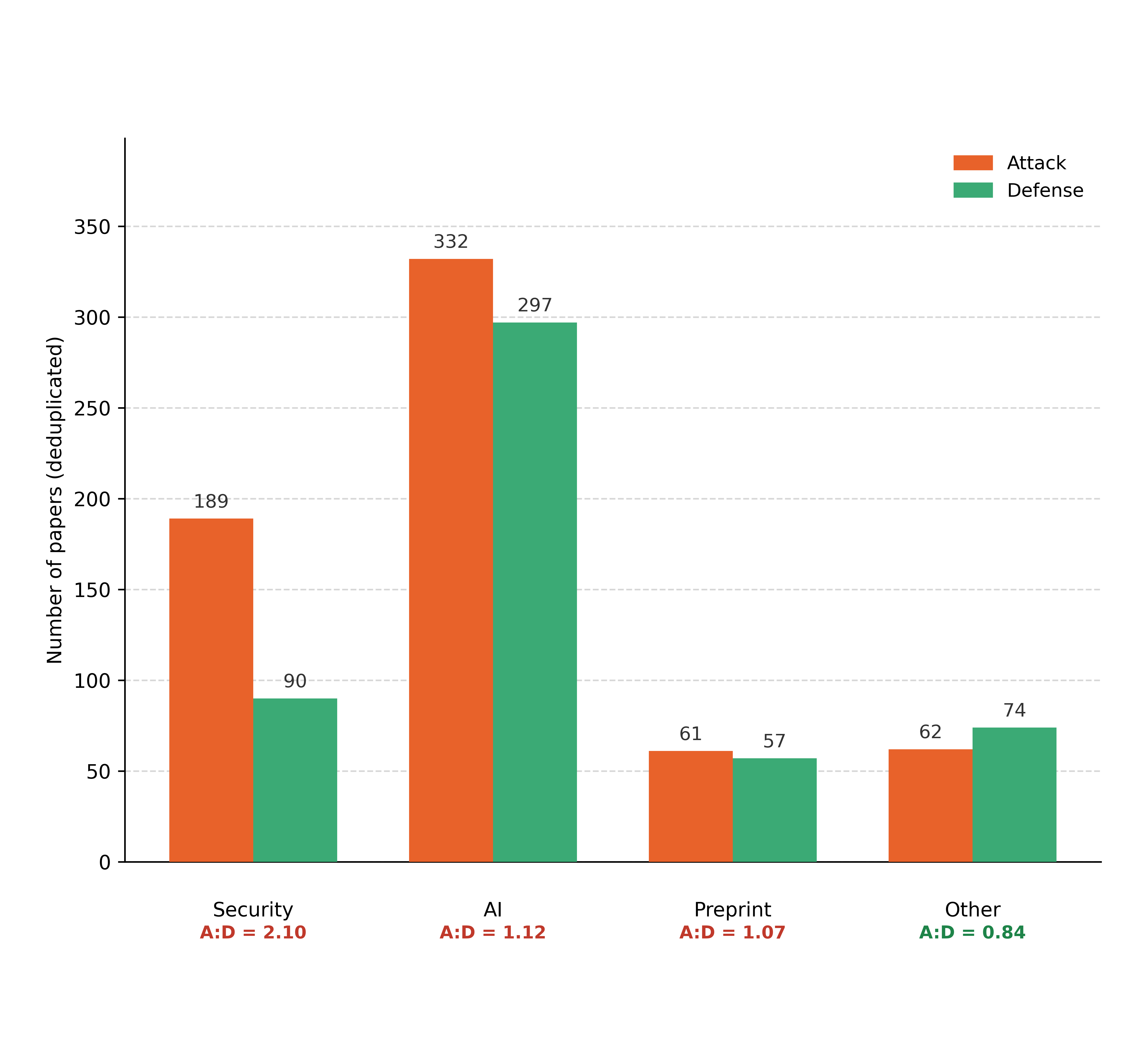}
        \caption{}
        \label{fig:venue_category}
    \end{subfigure}
    \hfill
    \begin{subfigure}[t]{0.49\linewidth}
        \centering
        \includegraphics[width=\linewidth]{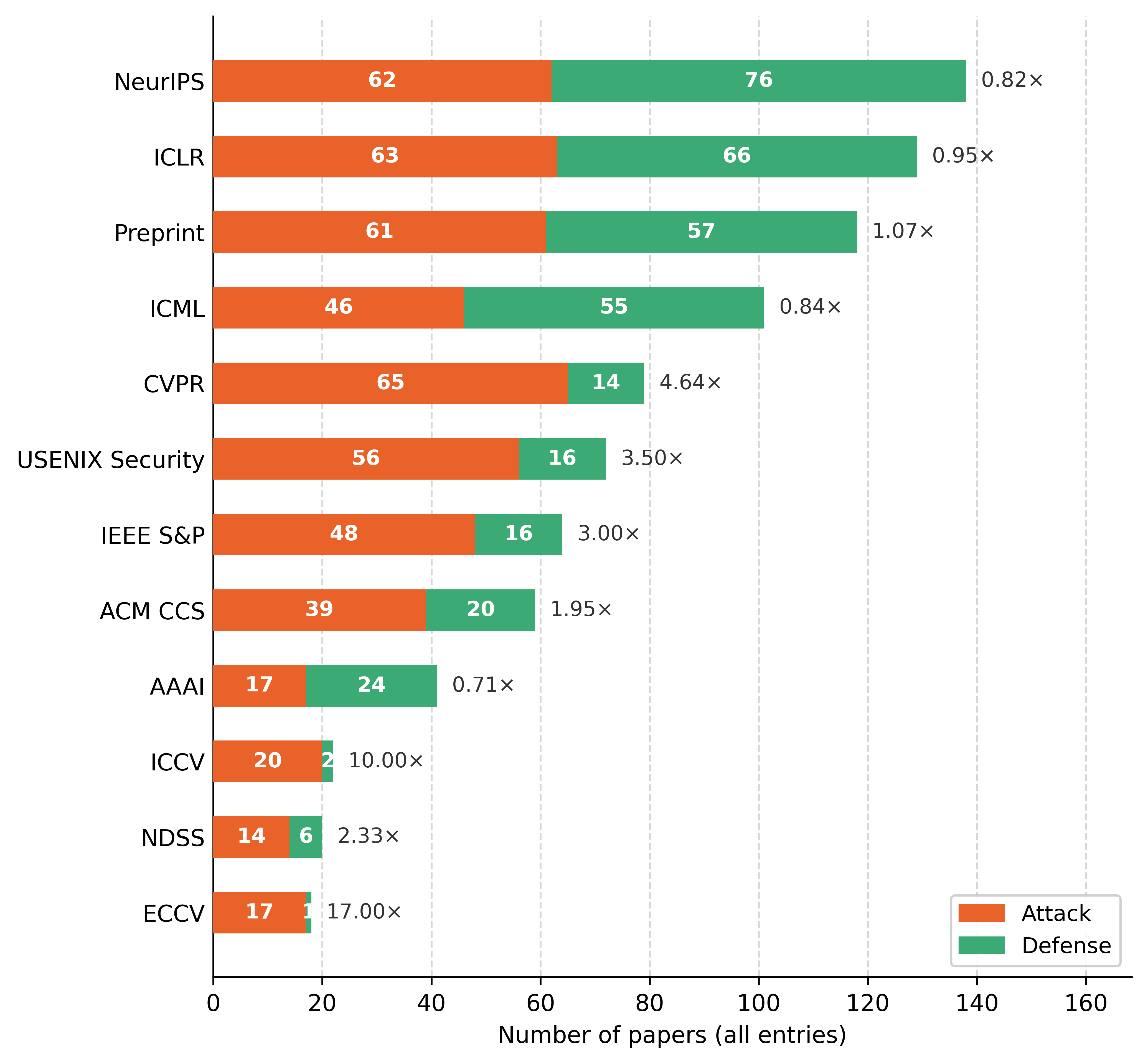}
        \caption{}
        \label{fig:top12_venues}
    \end{subfigure}
    \caption{Venue distribution of attack vs.\ defense papers. (a) By venue category; (b) Top 12 individual venues.}
    \label{fig:venue_distribution}
\end{figure}

Figure~\ref{fig:venue_distribution} shows that the attack--defense asymmetry is also visible at the venue level in the SoK corpus.
As shown in Figure~\ref{fig:venue_distribution}(a), security venues publish 189 attack papers and 90 defense papers, corresponding to an attack-to-defense ratio of 2.10:1. By contrast, AI venues are much closer to parity, with 332 attack papers and 297 defense papers, or an attack-to-defense ratio of 1.12:1. This contrast suggests that the imbalance is not simply a property of the topic itself, but may be also shaped by different publication venues. In particular, venues centered on security appear to lean toward attack-oriented contributions, whereas general AI venues appear relatively more open to defense-focused work.

This pattern becomes even clearer when examining individual flagship conferences. The four major security venues---USENIX Security (3.50$\times$), IEEE S\&P (3.00$\times$), NDSS (2.33$\times$), and ACM CCS (1.95$\times$)---all show an attack skew. Such a consistency suggests that the result is unlikely to be explained by the reviewing culture of any single conference; rather, it points to a broader publication norm within security research, i.e., attacks are favored. 
By contrast, several leading AI conferences show the opposite tendency: NeurIPS (0.82$\times$), AAAI (0.71$\times$), ICLR (0.95$\times$), and ICML (0.84$\times$) each include more defense papers than attack papers.

A further noteworthy pattern emerges within the AI category itself. Vision venues are attack-skewed: CVPR has an attack-to-defense ratio of 4.64$\times$, ICCV 10.00$\times$, and ECCV 17.00$\times$. This likely reflects the long-standing influence of the adversarial-examples literature in computer vision, where attacks became a central way to expose model brittleness and benchmark robustness, while defenses often struggled to establish lasting credibility. 

The contrast is therefore analytically informative: the relative balance observed in the aggregate AI category masks important differences across sub-fields. 
Some areas, especially computer vision, remain strongly attack-centered, while others are considerably more receptive to defense research. 
This suggests that the attack--defense imbalance is shaped not only by the technical characteristics of different research areas, but also by the norms and evaluation criteria of different research communities and publication venues.

\subsection{Trend Over Time}

\begin{figure}[t]
    \centering
    \includegraphics[width=\linewidth]{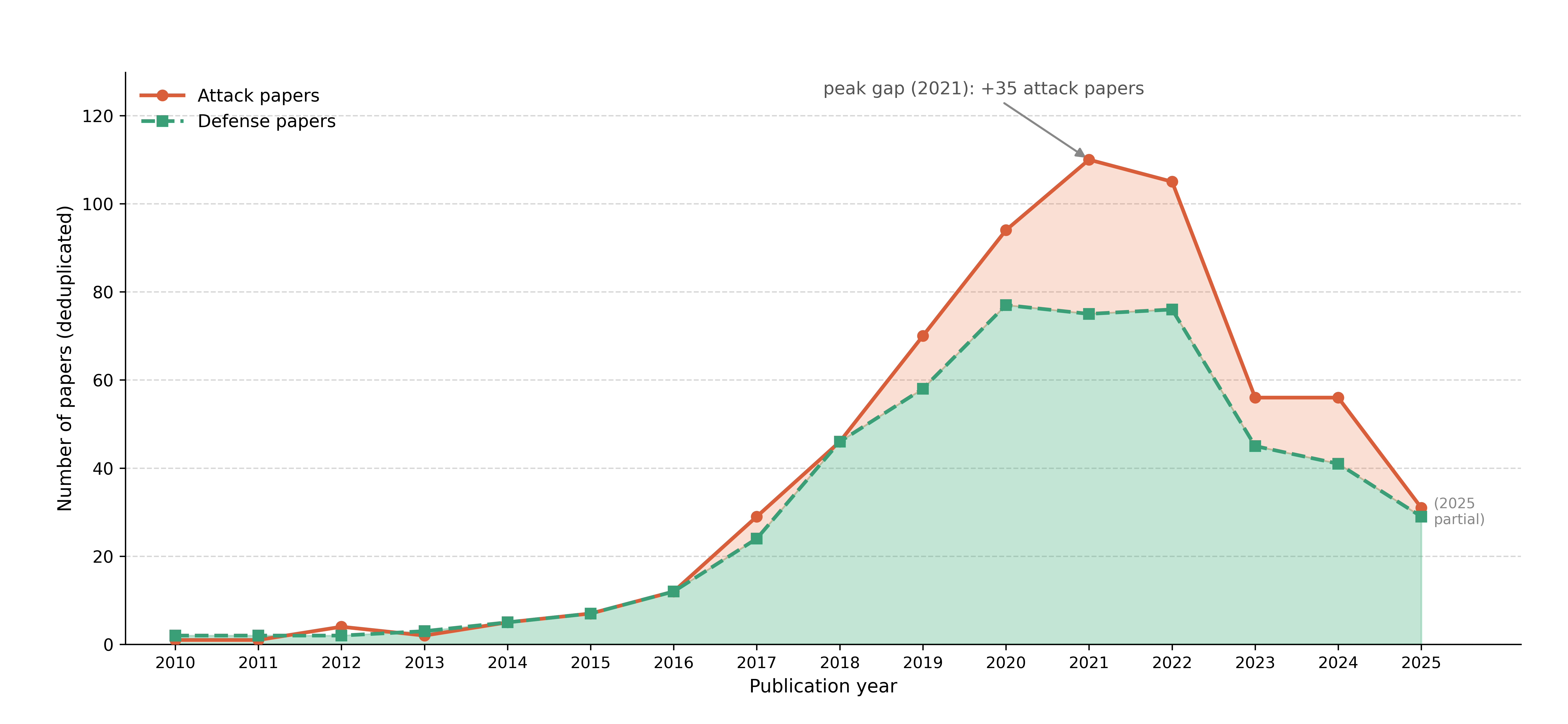}
    \caption{Attack vs. defense paper counts over time (2010–2025).}
    \label{fig:year_trend}
\end{figure}

As shown in Figure~\ref{fig:year_trend}, attack and defense papers track each other relatively closely in the earliest years of the SoK corpus, with both remaining at low levels through approximately 2015. Around 2016, however, the two trajectories began to diverge. From that point onward, attack papers outnumber defense papers represented in the corpus, and the gap becomes persistent rather than occasional. 
It suggests that the imbalance is not confined to a single mature phase of the field; rather, it emerges alongside the field's broader expansion and remains embedded as the literature grows~\cite{dblp26publications}.

The widening from 2017 to 2022 is  notable, where the largest single-year gap appears in 2021.
This period may closely align with the transformation of AI security from a relatively specialized niche into a more established and rapidly growing subfield, marked by the emergence of benchmark datasets, more standardized threat models, dedicated workshops, and increasing visibility at flagship conferences. 
However, the apparent narrowing after 2022 should be interpreted with caution. 
Because the corpus is constructed from cited studies in SoK papers, it is subject to citation lag: recent work is inevitably underrepresented, since SoKs require time to be researched, written, reviewed, and published. As a result, the decline in both series after 2022 is more plausibly understood as a sampling artifact than as evidence of a genuine slowdown in the literature or a substantive convergence between attack and defense output. 

\subsection{Other Survey Papers}

Recall that we also extended our analysis beyond SoKs to cited attack/defense research in another 21 survey papers. 
Among these, all 21 surveys cited at least one attack paper, while 19 cited at least one defense paper. 
A valid attack-to-defense ratio could be computed for 15 of the 21 surveys. 
For this extended sample, we calculated the mean, standard deviation, minimum, and maximum for the number of attack papers, the number of defense papers, and the attack-to-defense ratio, as shown in Table~\ref{tab:stats_mwu}. 
At the descriptive level, the mean number of attack papers exceeds the mean number of defense papers.

\begin{table}[ht]
\centering
\caption{Statistics of Attack and Defense Paper Counts in Survey Papers}
\label{tab:stats_mwu}
\begin{tabular}{lrrrrr}
\hline
\textbf{Metric} & \textbf{N} & \textbf{Mean} & \textbf{Std} & \textbf{Min} & \textbf{Max} \\
\hline
Count of Attack Papers  & 21 & 41.81 & 27.98 & 14  & 110   \\
Count of Defense Papers & 19 & 29.32 & 16.42 &  6  &  68   \\
Attack-to-Defense Ratio   & 15 &  2.153 &  1.240 & 0.315 & 5.125 \\
\hline
\end{tabular}
\end{table}





%% file: dilemma_for_defense_research.tex
\section{Dilemmas for Defense Research}

\subsection{Novelty Asymmetry}

A first dilemma is the asymmetry in how novelty is recognized and rewarded. Attack papers can often establish a contribution by identifying a new attack vector, applying a known technique to a new target, or showing that an existing system fails under a previously untested threat model. In fast-moving areas such as LLM security, this creates a particularly fertile publication environment for attacks, because each newly deployed model or interface generates another attack surface. Meeus et al. explicitly describe ``a surge in research on MIAs against LLMs'' following the release of GPT-family models, while also noting that many such papers rely on weak or flawed evaluation practices~\cite{meeus2025sok}. The broader implication is that the threshold for novelty is structurally lower on the attack side: demonstrating one new failure mode is often sufficient to claim a publishable contribution, whereas demonstrating one defense that remains effective under adaptive evaluation is a substantially more demanding bar. As a result, the incentive system tends to generate attack proliferation more easily than defense consolidation.

\subsection{Evaluation Asymmetry}

A second dilemma is evaluation asymmetry. Attacks and defenses are not judged by symmetrical standards. An attack can succeed by breaking one version of one model under favorable conditions; a defense is expected to remain effective across many models, datasets, and attacker adaptations, all while preserving utility, efficiency, and sometimes fairness. This asymmetry is documented directly in multiple SoKs. Suya et al. identify recurring problems such as non-adaptive evaluation, cherry-picked threat models, and failure to test against sufficiently strong adversaries~\cite{suya2024sok}. Meeus et al. make a similar point in the LLM privacy context, where many attack papers claim strong results despite weak methodological controls~\cite{meeus2025sok}. Carletti et al.'s systematization of gradient inversion attacks adds another concrete example: attacks are often strongest in highly favorable conditions, such as batch size \(= 1\), while defenses must prove themselves across broader and more realistic settings~\cite{carletti2025sok}. The consequence is structural. Defenses are penalized for incompleteness in a way attacks are not, because defensive claims are implicitly universal, whereas attack claims need only be existential.

\subsection{Publication Asymmetry}

Publication incentives likely further amplify this asymmetry. A surprising vulnerability (in attack research) is often easier to present and evaluate as a research contribution than incremental hardening (in defense research). 
Attack papers typically follow a clear and compelling structure: a trusted system is deployed, a key assumption is shown to fail, and the paper exposes a previously unrecognized weakness. 
Defense papers, by contrast, often depend on a more fragile form of claim, namely, that a method withstands all evaluated attacks or mitigates a particular class of failures under specified assumptions. Such claims are inherently more difficult to validate and are more susceptible to later reversal if the defense is subsequently broken. This helps explain why defense papers may appear less publishable even when they are technically substantive.
Duddu et al. provide indirect support for this interpretation by showing that defenses can conflict with one another across threat categories, such that a method that improves robustness in one dimension may simultaneously weaken privacy or security in another~\cite{duddu2024sok}. If defense mechanisms must be assessed not only on whether they are effective in isolation, but also on whether they introduce harmful trade-offs with other protections, then the evidentiary burden for publishable defense research becomes even greater.

\subsection{Arms-Race Dynamics}

Another likely driver is the arms-race structure of the field itself. 
Multiple SoKs describe domains in which attacks and defenses co-evolve, but not with equal reputational standing. Li et al. identify a related dynamic in certified robustness: methods that offer meaningful guarantees for smaller models often do not scale cleanly to larger or more realistic settings, thereby necessitating repeated rounds of certification research~\cite{li2023sok}. In such environments, defenses can appear provisional by construction. Even when effective, they are often understood as temporary responses within an ongoing contest rather than as stable solutions.
Attack papers, by contrast, may retain their prestige more easily because the discovery of a new vulnerability is often treated as a foundational insight into the behavior or limitations of a system, even if the specific exploit later becomes outdated. This asymmetry in perceived durability may sustain the broader status hierarchy of the field: attacks are often viewed as uncovering fundamental truths about systems, whereas defenses are more often treated as contingent efforts to repair or contain those weaknesses.

\subsection{Cost and Complexity Asymmetry}

A further reason defenses may lag is that attacks and defenses often differ sharply in implementation cost and organizational scope. Horv\'ath et al. show that physical side-channel attacks on neural networks can be mounted in laboratory settings using relatively accessible measurement setups and off-the-shelf hardware~\cite{horvath2024sok}. By contrast, effective defenses may require secure hardware redesigns, masking strategies, architectural modifications, or noise-injection schemes that must be integrated into production systems. Similarly, Du et al. observe that some gradient inversion attacks can be mitigated by comparatively simple software interventions such as quantization or sparsification, but doing so without unacceptable loss in model utility requires careful tuning and deployment-specific engineering~\cite{du2025sok}. In other words, proving that a system can be attacked is often an academic problem, while making it robust can become an industrial systems problem. This division likely skews the literature toward attack papers because academia is structurally better positioned to demonstrate vulnerabilities than to ship and maintain production-quality protections.

\subsection{The Industry--Academia Gap}

The visibility of attacks may also be amplified by an industry--academia gap in what gets published. Nayan et al. note that, for several model-extraction threat categories, they were ``not aware of industry defense solutions''~\cite{nayan2024sok}. This statement is revealing not because it proves defenses do not exist, but because it highlights how little of industrial defensive practice enters the public record. Companies have strong incentives to protect deployed models, but those protections may be proprietary, rather than disclosed in academic venues. Attack papers face the opposite publication dynamic: they are often publishable precisely because they reveal external vulnerabilities, and companies have limited incentive to publicize their mitigations in comparable detail, possibly resulting in a systematic observational bias. The academic literature can become attack-rich and defense-poor even if some defenses are actively used in practice, simply because offensive findings are more visible than defensive operations.


%% file: alternative_views.tex
\section{Alternative Views}

The argument of this paper is not that attack research is undesirable, nor that the literature should contain fewer attack papers. 
Rather, conditional on the field's stated goal of improving AI security, the current incentive structure appears to direct relatively less attention toward defense research than toward attack research.
A meaningful counterposition must therefore do more than defend the value of attack research. 
Instead, it must challenge the inference that \textbf{AI security research should better incentivize defense research}, or at least challenge the view that the observed publication imbalance constitutes evidence of a problem. 
Below, we consider several alternative views that more directly address that claim.

\subsection{The Current Level of Defense Research Is Already Appropriate}

\textbf{Counterposition.} The attack--defense imbalance may not indicate a problem. It may instead reflect what the field most needs at its current stage. If researchers still do not fully understand the main threats, attack capabilities, and failure modes of AI systems, then it can be reasonable for attack research to receive more attention than defense research. On this view, producing more defense papers would not necessarily improve security. It could instead lead to defenses that are premature, weak, or misleading, because the underlying threat landscape has not yet been fully mapped.

This is a strong alternative to the argument of this paper. Its core idea is simple: before a field can reliably defend systems, it must first understand how those systems fail. Attack papers can help do that. They expose hidden assumptions, reveal unexpected weaknesses, and clarify what kinds of threats defenses will eventually need to address. If this basic mapping work is still incomplete, then an attack-heavy literature may be appropriate rather than problematic.

The SoKs provide some support for this view. Abdullah et al. map the adversarial machine learning threat landscape and clarify the range of attack surfaces that defenses would need to cover~\cite{abdullah2021sok}. Li et al.'s discussion of certified robustness likewise depends on a prior body of attack research against which defensive guarantees can be defined and tested~\cite{li2023sok}. Du et al.'s survey of gradient inversion attacks further shows that attack-focused work can also correct mistaken beliefs by showing that some threats are weaker than earlier papers suggested~\cite{du2025sok}. 
These examples support the idea that attack research can be necessary not only for finding problems, but also for understanding which problems actually matter.

Even so, this counterposition is less convincing if it is applied to the entire field without qualification. In several areas, including adversarial examples, membership inference, and model extraction, researchers have already spent years identifying attacks and organizing them into clearer threat categories. 
In such cases, it becomes harder to argue that the field is still mainly in an early discovery phase. Once a threat area is already well mapped, the value of yet another attack paper may be lower than the value of stronger mitigation.

The temporal pattern in this paper also raises doubts about this counterposition. If the attack-heavy structure were mainly a temporary feature of early-stage exploration, then one might expect the gap between attack and defense papers to narrow as the field matures. But the evidence here suggests the opposite: the gap persisted and even widened during the field's main growth period. That pattern is harder to explain as a temporary need for threat discovery alone.

A more moderate version of the counterposition is therefore more plausible. Some subfields may still reasonably require attack-heavy research, while others may now need more high-quality defense work. On this interpretation, the problem is not that every area of AI security should immediately shift toward defense, but that the field as a whole may still reward attack research too strongly even in domains where the central challenge has already shifted from discovering threats to mitigating them.

\subsection{What Is Needed Is Not More Defense Papers, but Better Translation}

\textbf{Counterposition.} The main problem may not be that the field produces too little defense research. The real problem may be that existing defenses are hard to turn into real protection. On this view, asking for more defense papers points to the wrong bottleneck. What is needed is not simply more published defenses, but better translation of defensive ideas into systems that organizations can actually use.

This is a serious alternative to the paper's argument because it challenges a possible remedy---more defense research. Even if attack papers clearly outnumber defense papers, that does not by itself show that the field mainly needs more defense research. In many cases, defenses do not fail because no one has proposed them, but because they are difficult to integrate into real systems, reduce performance, add latency or cost, or require changes to hardware, infrastructure, or workflows. If that is the real obstacle, then producing additional defense papers may do less for security than work focused on deployment, maintenance, and organizational adoption.

The SoKs provide support for this view. As mentioned previously, Horv\'ath et al. show that defending against physical side-channel attacks may require hardware redesign and other interventions that go far beyond what a typical academic defense paper can offer~\cite{horvath2024sok}. Du et al. show that some mitigations can reduce gradient inversion risk, but making those mitigations practical without seriously harming model quality is largely an engineering problem~\cite{du2025sok}. Nayan et al. likewise suggest that some useful defenses against model extraction may already exist in industry, even if they do not appear in the academic literature in standard paper form~\cite{nayan2024sok}. On this view, defense may look scarce in publications, not because too little thought has gone into it, but because much of the most important work happens outside the kinds of outputs that paper counts can easily capture.

This counterposition may be correct in some domains. If so, then the practical claim of our paper should be broader than a call for more attention to defense research alone. What may be needed is a greater investment in defensive capacity more generally: deployment engineering, hardening informed by red-team results, maintaining benchmarks, secure hardware, and better pathways for industry to share defensive lessons.

Still, this counterposition does not fully overturn the original argument. If public research produces too little open and scrutinizable defense knowledge, then better translation by itself is not enough. Defenses that exist only as private engineering practices do not give the field a shared and cumulative understanding of what works. A better conclusion, then, may be that the bottleneck is partly translational, but that this is an additional reason, not a reason to stop, for improving incentives around public defense work.

%% file: limitations.tex
\section{Limitations}

We note that the evidence base for this analysis consists of cited papers from 16 SoKs and 21 surveys. 
While these papers provide broad and valuable coverage of AI security, they may not fully capture some broader topics, such as safety, alignment, governance, and trustworthiness. 
A more comprehensive analysis would benefit from incorporating a larger independent corpus, for example by drawing from sources such as DBLP and conducting topic classification at scale.
Still, our current pipeline offers a practical and well-motivated ``first-step'' assessment. 
Since the unit of analysis is based on citations within SoKs and surveys, the resulting picture can reflect how these papers curate and represent the field, rather than a direct measure of the field’s total research output. 
Future work could further strengthen by triangulating SoK- and survey-based findings with a much broader literature corpus.

Another limitation of this work is that, while we argue that defense research may be under-incentivized relative to attack research, we do not specify which interventions would mitigate this imbalance. We view the design of effective incentive mechanisms for defense research as an important open question for future work.

%% file: conclusion.tex
\section{Conclusion}

Our analysis suggests that the imbalance between attack and defense research in AI security is not just a matter of paper counts. 
It appears to reflect a broader incentive structure within the field. 
Across the SoKs and surveys we examined, attack work is often more numerous and more publicly rewarded, while defense work is often expected to meet a higher standard of robustness, realism, or generality before it is treated as a strong contribution. 
We do not argue that attack research should be reduced. Attack research remains essential for identifying threats, testing assumptions, and clarifying what defenses must address. Nor does our argument imply that the only missing input is a larger number of defense papers. In many domains, the bottleneck may also lie in translation: engineering, deployment, maintenance, organizational adoption, and other forms of defensive capacity that are not well captured by standard publication metrics. 
AI security should better incentivize defensive capacity in a broader sense: stronger public defense research, more realistic defense evaluation, clearer pathways from defensive ideas to deployment, and more institutional support for mitigation-oriented work. A mature security field should not only be good at showing how systems fail. It should also be good at building, validating, and sharing ways to make them more secure and much safer.

%% file: appedix_soks.tex
\section{SoKs}
\label{app:soks}

\begin{table*}[h]
\centering
\caption{Overview of SoK papers about security and privacy, sorted by venue and year.
         \#A and \#D denote the number of attack papers and defense papers cited,
         respectively.}
\label{tab:sok_overview}
\small
\begin{tabular}{@{}
  >{\raggedright\arraybackslash}p{1.5cm}
  >{\raggedright\arraybackslash}p{1.4cm}
  >{\raggedright\arraybackslash}p{0.7cm}
  >{\raggedright\arraybackslash}p{3.8cm}
  >{\raggedright\arraybackslash}p{0.6cm}
  >{\raggedright\arraybackslash}p{0.6cm}
  >{\raggedright\arraybackslash}p{0.9cm}
@{}}
\toprule
\textbf{Paper} & \textbf{Venue} & \textbf{Year} & \textbf{Title} & \textbf{\#A} & \textbf{\#D} & \textbf{A:D} \\
\midrule
\cite{papernot2018sok}   & IEEE\newline EuroS\&P & 2018 & SoK: Security and Privacy in Machine Learning                                           & 40  & 22  & 1.8  \\
\cite{abdullah2021sok}   & IEEE\newline S\&P     & 2021 & SoK: The Faults in Our ASRs: An Overview of Attacks against ASR and Speaker ID Systems  & 39  & 20  & 2.0  \\
\cite{li2023sok}         & IEEE\newline S\&P     & 2023 & SoK: Certified Robustness for Deep Neural Networks                                      & 19  & 162 & 0.1  \\
\cite{salem2023sok}      & IEEE\newline S\&P     & 2023 & SoK: Let the Privacy Games Begin! A Unified Treatment of Data Inference Privacy in ML   & 45  & 6   & 7.5  \\
\cite{wenger2023sok}     & IEEE\newline S\&P     & 2023 & SoK: Anti-Facial Recognition Technology                                                 & 48  & 13  & 3.7  \\
\cite{duddu2024sok}      & IEEE\newline S\&P     & 2024 & SoK: Unintended Interactions among Machine Learning Defenses and Risks                  & 78  & 60  & 1.3  \\
\cite{noppel2024sok}     & IEEE\newline S\&P     & 2024 & SoK: Explainable Machine Learning in Adversarial Environments                           & 50  & 68  & 0.7  \\
\cite{khanna2024sok}     & IEEE\newline SaTML    & 2024 & SoK: A Review of Differentially Private Linear Models for High-Dimensional Data         & 0   & 31  & 0.0  \\
\cite{suya2024sok}       & IEEE\newline SaTML    & 2024 & SoK: Pitfalls in Evaluating Black-Box Attacks                                           & 172 & 5   & 34.4 \\
\cite{meeus2025sok}      & IEEE\newline SaTML    & 2025 & SoK: Membership Inference Attacks on LLMs are Rushing Nowhere (and How to Fix It)      & 43  & 8   & 5.4  \\
\cite{horvath2024sok}    & USENIX\newline Sec.   & 2024 & SoK: Neural Network Extraction through Physical Side Channels                           & 70  & 16  & 4.4  \\
\cite{nayan2024sok}      & USENIX\newline Sec.   & 2024 & SoK: All You Need to Know about On-Device ML Model Extraction                          & 30  & 26  & 1.2  \\
\cite{carletti2025sok}   & USENIX\newline Sec.   & 2025 & SoK: Gradient Inversion Attacks in Federated Learning                                   & 50  & 28  & 1.8  \\
\cite{du2025sok}         & USENIX\newline Sec.   & 2025 & SoK: On Gradient Leakage in Federated Learning                                          & 40  & 14  & 2.9  \\
\cite{nowmi2025critical} & arXiv                 & 2025 & Critical Evaluation of Quantum Machine Learning for Adversarial Robustness              & 28  & 26  & 1.1  \\
\cite{wang2025sok}       & arXiv                 & 2025 & SoK: Evaluating Jailbreak Guardrails for Large Language Models                          & 35  & 58  & 0.6  \\
\bottomrule
\end{tabular}
\end{table*}

%% file: appendix_surveys.tex
\section{Surveys}
\label{app:surveys}

\begin{table*}[h]
\centering
\caption{Overview of survey papers on AI security and privacy, sorted by type and year.
         \#A and \#D denote the number of directly related attack and defense papers cited by each survey,
         respectively. ``---'' indicates data not reported.}
\label{tab:survey_type_overview}
\small
\begin{tabular}{@{}
  >{\raggedright\arraybackslash}p{2cm}
  >{\raggedright\arraybackslash}p{2cm}
  >{\centering\arraybackslash}p{0.8cm}
  >{\centering\arraybackslash}p{0.8cm}
  >{\centering\arraybackslash}p{0.8cm}
  >{\centering\arraybackslash}p{1.2cm}
@{}}
\toprule
\textbf{Type} & \textbf{Paper} & \textbf{Year} & \textbf{\#A} & \textbf{\#D} & \textbf{A:D} \\
\midrule
\multirow{5}{=}{Adversarial Samples}
  & \cite{serban2020adversarial} & 2020 & 27  & 45  & 0.60 \\
  & \cite{nguyen2023physical}    & 2023 & 25  & 7   & 3.57 \\
  & \cite{guesmi2023physical}    & 2023 & 94  & --- & ---  \\
  & \cite{wei2024physical}       & 2024 & 86  & --- & ---  \\
  & \cite{wei2026visual}         & 2026 & 56  & 31  & 1.81 \\
\midrule
\multirow{3}{=}{Deepfake}
  & \cite{edwards2024review} & 2024 & --- & 29  & ---  \\
  & \cite{pei2024deepfake}   & 2024 & 110 & 30  & 3.67 \\
  & \cite{li2025survey}      & 2025 & --- & 46  & ---  \\
\midrule
\multirow{2}{=}{Federated Learning}
  & \cite{yang2023gradient}   & 2023 & 14  & --- & ---  \\
  & \cite{zhao2025federation} & 2025 & 23  & 13  & 1.77 \\
\midrule
\multirow{4}{=}{LLM}
  & \cite{yi2024jailbreak}       & 2024 & 53 & 30 & 1.77 \\
  & \cite{dong2025safeguarding}  & 2025 & 37 & 17 & 2.18 \\
  & \cite{knowlton2026prompt}    & 2026 & 25 & 31 & 0.81 \\
  & \cite{hakim2026jailbreaking} & 2026 & 41 & 8  & 5.13 \\
\midrule
\multirow{3}{=}{Machine Unlearning}
  & \cite{shaik2024exploring} & 2024 & --- & 40  & ---  \\
  & \cite{liu2025threats}     & 2025 & 20  & 6   & 3.33 \\
  & \cite{nguyen2025survey}   & 2025 & 14  & 68  & ---  \\
\midrule
\multirow{4}{=}{MIA}
  & \cite{hu2022membership}  & 2022 & 64 & 35  & 1.83 \\
  & \cite{hu2023defenses}    & 2023 & 17 & 54  & 0.31 \\
  & \cite{bai2024membership} & 2024 & 24 & 15  & 1.60 \\
  & \cite{wu2025membership}  & 2025 & 25 & --- & ---  \\
\midrule
\multirow{2}{=}{Model Stealing}
  & \cite{gencc2023taxonomic} & 2023 & 18 & --- & ---  \\
  & \cite{oliynyk2023know}    & 2023 & 76 & 36  & 2.11 \\
\midrule
Speech & \cite{tan2022adversarial} & 2022 & 29 & 16 & 1.81 \\
\bottomrule
\end{tabular}
\end{table*}